\documentclass{article}
\usepackage{spconf,amsmath,graphicx}

\usepackage{amsfonts}
\usepackage{multirow}
\usepackage{xcolor}
\usepackage{hyperref}
\usepackage{nicefrac}
\usepackage{booktabs}
\usepackage{makecell}
\usepackage{multirow}

\def\tablespace{\vspace{9pt}}

\newcommand{\newpara}[1]{\vspace{11pt}\noindent\textbf{#1}}

\title{Look who's not talking}
\name{Youngki Kwon$^{1}$, Hee Soo Heo$^{1}$, Jaesung Huh$^{2}$, Bong-Jin Lee$^{1}$, Joon Son Chung$^{1}$}
\address{$^{1}$Naver Corporation, South Korea\\ 
$^{2}$Visual Geometry Group, University of Oxford, UK}

\begin{document}

\maketitle

\begin{abstract}
The objective of this work is speaker diarisation of speech recordings ‘in the wild'. The ability to determine speech segments is a crucial part of diarisation systems, accounting for a large proportion of errors. In this paper, we present a simple but effective solution for speech activity detection based on the speaker embeddings. In particular, we discover that the {\em norm} of the speaker embedding is an extremely effective indicator of speech activity. The method does not require an independent model for speech activity detection, therefore allows speaker diarisation to be performed using a unified representation for both speaker modelling and speech activity detection. We perform a number of experiments on in-house and public datasets, in which our method outperforms popular baselines.
\end{abstract}

\vspace{10pt}

\section{Introduction}

Automatic transcription of meetings and conversations is a very attractive ability, since it enables human communications to be archived in a machine indexable format. While many researchers in the speech domain focus on the speech-to-text part of the problem, being able to determine ‘who said when' is an equally important and challenging task. 
While the training of speech and speaker recognition involves classifying the input into a fixed number of classes (such as phonemes and speakers), speaker diarisation requires assigning parts of speech to an unknown number of classes, if any.

The challenging problem has been the subject of active research over the recent years. The majority of works in speaker diarisation use at least two independent components -- a speech activity detector to determine whether somebody is speaking at a particular time, and a speaker embedding extractor to determine who is speaking if somebody is speaking. 

Speech activity detection (also known as voice activity detection) is an important front-end functionality in speech processing. While there are many works in this task using both feature engineering~\cite{nemer2001robust, wu2005robust, tanyer2000voice} and deep learning~\cite{eyben2013real, chang2018temporal}, many researchers in diarisation have used off-the-shelf packages~\cite{johnston2012webrtc, cernak2017bob}, while a few have trained deep neural networks specifically on task-specific datasets~\cite{sun2018speaker,sell2018diarization,park2019second}.

A speaker embedding model is the second key component of most diarisation systems. Traditionally, speaker models are constructed with Gaussian mixture models (GMMs) and i-vectors~\cite{dehak2010front,cumani2013probabilistic,matvejka2011full}, but more recently deep learning has been proven effective for speaker modelling~\cite{variani2014deep,ghalehjegh2015deep,snyder2017deep,snyder2018x}. Typical diarisation systems are based on the clustering of these speaker embeddings using a range of algorithms such as spectral clustering~\cite{ning2006spectral,wang2018speaker}, agglomerative hierachical clustering~\cite{garcia2017speaker,maciejewski2018characterizing,sell2014speaker,sun2018speaker,sell2018diarization} and k-means clustering~\cite{dimitriadis2017developing,wang2018speaker}.
While there are some recent works on fully end-to-end speaker diarisation, the experiments are constrained to limited conditions such as telephone speech~\cite{zhang2019fully,fujita2019end,fujita2019end2}.

The traditional pipeline requires at least two models and feature extraction steps as described above.
This raises an interesting question: ‘‘Can we also use the speaker embeddings to obtain speech activity labels?''. We might expect the answer to be ‘no', since the speaker networks are only trained to discriminate between speakers.
However, since the speaker embeddings are able to discriminate one person’s speech from another, it might also be able to discriminate speech from non-speech. In this paper, we investigate the use of speaker recognition features for speech activity detection. 
It has been suggested that the norm of an image embedding is related to the ability to discriminate the object in a target domain~\cite{taigman2015web,subramanya2017confidence}, or in other words, the {\em confidence}. Since speaker recognition networks are trained on speech segments, the network should not be confident for a non-speech input since it is unable to detect target patterns. Therefore, we use the norm of the embedding as an indicator of speech activity, since this value is correlated with speaker confidence of each frame. This strategy only requires one forward pass through the feature extractor, which reduces computational requirements compared to the previous two-stage pipeline.
Additionally, we can expect a lower speaker confusion error by selecting more confident frames for speaker recognition using the feature extractor.

We perform a range of experiments on an unreleased internal dataset of real-world meetings, as well as the public VoxConverse and DIHARD datasets. We observe that speaker embeddings trained on the verification task work surprisingly well for speech activity detection, even though the embedding extractor never has never seen non-speech during training. Our method demonstrates competitive performance against popular baselines across a wide range of experiments.


\section{Methods}
\label{sec:methods}

This section describes the core methods and architecture used in the diarisation system.

\subsection{Speaker representations}

Obtaining good speaker representations is at the heart of the diarisation problem. In the following paragraphs, we describe the method for training and extracting speaker embeddings using a deep neural network.

\newpara{Input representations.}
We extract spectograms from each utterance with a hamming window of 25 millisecond width and 10 millisecond stride. 
64-dimensional Mel filterbanks are used as the input to the network.
Mean and variance normalisation (MVN) is performed on every frequency bin of the spectrogram and filterbank at utterance-level using instance normalisation~\cite{ulyanov2016instance}.

\newpara{Embedding extractor.}
Residual networks~\cite{He16} have been used successfully in speaker recognition~\cite{Xie19a,cai2018exploring,Chung18a,chung2020defence}. The basic architecture is the same as the ResNet-34 network described in~\cite{He16}, except with pre-activation residual units~\cite{he2016identity}. The network architecture is given in Table~\ref{table:convnet}. 

\begin{table}[ht]
\renewcommand\arraystretch{1.1}
\centering
\caption{The ResNet-34 architecture. }
\label{table:convnet}
\tablespace
\begin{tabular}{ c|c|c }
\textbf{Layer}   & \textbf{\# filts.} & \textbf{Outputs}  \\ \hline
 & \makecell{$7 \times 7,64$, stride $2 \times 1$ \\ $ 3 \times 3$, Maxpool, stride $1 \times 1$}  & \makecell{$32 \times T \times 64$} \\   \hline
block1  & $\begin{bmatrix}  3 \times 3, 64 \\ 3 \times 3, 64 \end{bmatrix} \times 3 $, stride 1 & $32 \times T \times 64$        \\\hline
block2 & $\begin{bmatrix}  3 \times 3,128 \\ 3 \times 3,128 \end{bmatrix} \times 4 $, stride 2 & $16 \times \nicefrac{T}{2} \times 128$    \\\hline
block3 & $\begin{bmatrix}  3 \times 3,256 \\ 3 \times 3,256 \end{bmatrix} \times 6 $, stride 2 & $8 \times \nicefrac{T}{4} \times 256$      \\\hline
block4  & $\begin{bmatrix}  3 \times 3,512 \\ 3 \times 3,512 \end{bmatrix} \times 3 $, stride 2 & $4 \times \nicefrac{T}{8} \times 512$      \\\hline
\end{tabular}
\end{table}

The output from the embedding extractor is aggregated over time using a temporal average pooling (TAP) layer, then passed through a linear projection layer to obtain an utterance-level embedding.

\newpara{Objective function.}
We train the speaker embedding extractor using a combination of a classification loss and the hard negative mining loss, as proposed in~\cite{heo2019end}. 

The classification loss is a standard categorical cross-entropy loss, defined as:

\begin{equation}
\mathcal{L}_\text{CE}=-\frac{1}{N}\sum_{i=1}^N\log \frac{e^{\mathbf{W}^T_{y_i}\mathbf{x}_i+b_{y_i}}}{\sum_{j=1}^C e^{\mathbf{W}^T_{j}\mathbf{x}_i+b_{j}}}
\end{equation}  

The hard negative mining loss is defined as:

\begin{equation}
\mathcal{L}_\text{H}=\sum\limits_{i=1}^{N}\sum\limits_{\mathbf{W}_h \in \mathcal{H}_i}\log(1+\exp(\cos(\mathbf{W}_h,\mathbf{x}_i) - \cos(\mathbf{W}_{y_i},\mathbf{x}_i))),
\end{equation}

\noindent where $N$ is the batch size, $\mathbf{x}_i$ and $\mathbf{W}_{y_{i}}$ denote the embedding vector from the $i$'th utterance and the basis of the corresponding speaker, respectively, and $\mathcal{H}_i$ is the set of the top $\text{H}$ speaker bases with large $\cos(\mathbf{W}_h,\mathbf{x}_i)|_{h\neq y_{i}}$ values. The speaker basis for a particular speaker is one row vector of weight matrix of the output layer corresponding to the speaker. $\mathcal{H}_i$, the hard set for each sample, is selected for every minibatch based on cosine similarities between sample $\mathbf{x}_i$ and all speaker bases in the training set. Categorical cross entropy loss $\mathcal{L}_\text{CE}$ and $\mathcal{L}_\text{H}$ are combined with equal weights. 

\begin{equation}
\mathcal{L}_\text{overall}=  \mathcal{L}_\text{CE} + \mathcal{L}_\text{H} 
\end{equation}

\newpara{Training.}
The speaker embedding extractor is trained on the VoxCeleb2 dataset~\cite{Chung18a}.
During training, we use a fixed-length 2-second temporal segments, extracted randomly from each utterance.
The models are trained using four NVIDIA P40 GPUs each with 24GB memory for $100$ epochs, using a batch size of 800.
We use the Adam optimizer with an initial learning rate of $0.001$, and thereafter following a cosine annealing schedule~\cite{loshchilov2016sgdr}. 
The resultant model achieves 1.8\% equal error rate on the original VoxCeleb1 test set~\cite{Nagrani17}.

\subsection{Speech activity detection}

In previous diarisation systems, the speaker models are only used for representing the speaker information in the frames selected by the speech activity detection process. 
Since the speaker embeddings are able to discriminate one person’s speech from another, it should also be able to discriminate speech from non-speech.  

It has been suggested that the norm of an embedding is correlated to the confidence on the target task~\cite{taigman2015web}.
If the embedding vector is classified by a linear classifier such as an output layer activated by softmax function, the high value of norm means the large margin between embedding vector and hyper-plane~\cite{subramanya2017confidence}.

Since the network has been trained only on human speech, it is likely to exhibit very low confidence for non-speech inputs that has not been seen during training. In other words, we can use the speaker model for speech activity detection without any independent module or modification. We refer to our method as {\bf SpeakerNet SAD} in the results. 

In order to get fine-grained speech activity labels, we take every output from the embedding extractor and pass them through the projection layer without any temporal aggregation. This is in contrast to the use of the embeddings for speaker representations, where the embeddings are aggregated over 2-second windows using temporal average pooling (TAP).

\newpara{Computing the threshold.}
We propose two methods for tuning the threshold. The first is to set a manual threshold on the value of the norm using the development set, by performing experiments and finding the best results over a range of threshold values. We set a single threshold for all datasets described in Section~\ref{sec:exp}.

The second one is to estimate the optimal threshold for each utterance using a GMM model. In this approach, we train the GMM model with two mixture components to learn a distribution of the norm values from one utterance. The two mixtures in the model represent speech and non-speech clusters, respectively. After training the GMM model, we can estimate threshold $\mathcal{T}$ using following equation:
\begin{equation}
\mathcal{T}=\alpha \max\{\mu_0,\mu_1\}+(1-\alpha)\min\{\mu_0,\mu_1\},
\end{equation}
\noindent where $ \mu_0 $ and $ \mu_1$ are the mean values from two GMM mixtures, and $\alpha$ is the weight factor between two mean values. Since this approach enables the adapted estimation of the threshold for each utterance, we can set the thresholds which are robust across various dataset domains. In this research, we fix the value of $\alpha$ as 0.1 for all experiments without considering dataset domain.

\newpara{Post-processing.}
Based on the SAD results, we split each session into multiple speech activity segments. At this step, we apply end point detection (EPD) to compensate for excessively rapid changes in SAD results. We detect the start and end points by sliding a window of a certain size. In particular, the start point is identified as where a ratio of speech-activated frames exceed 70\%, and the end point is also identified following the same rule for non-speech frames. The EPD post-processing is applied to both the WebRTC-based baseline~\cite{johnston2012webrtc} and the proposed systems. Based on empirical results, we fix the sizes of sliding window to five (50 ms) and ten (100 ms) for the baseline and the proposed system, respectively.

\subsection{Speaker clustering}

We use the Agglomerative Hierarchical Clustering (AHC) to group the speaker embeddings \cite{day1984efficient}.
The AHC algorithm can cluster representations based on a distance threshold or the number of clusters, but in most previous works, a distance threshold is set manually on the development set.
Although this usually results in a reasonably good performance in the target domain, a threshold set on one dataset often does not generalise well to other datasets.
Since we perform experiments across a number of different domains, we automatically select the best number of clusters for each session or video based on the silhouette score~\cite{rousseeuw1987silhouettes} for $ 2 \leq C \leq 10$. 

The silhouette score is an interpretation of consistency within clusters of data, therefore can be seen as a measure of confidence. The silhouette score is defined in terms of the mean intra-cluster distance:
\begin{equation}
a(i)={\frac {1}{|C_{i}|-1}}\sum _{j\in C_{i},i\neq j}d(i,j)
\end{equation}

\noindent and the mean nearest-cluster distance:
\begin{equation}
b(i)=\min _{k\neq i}{\frac {1}{|C_{k}|}}\sum _{j\in C_{k}}d(i,j)
\end{equation}

\noindent for each sample. Specifically, the Silhouette score $s(i)$ for a sample is: \begin{equation}
s(i)={\frac {b(i)-a(i)}{\max\{a(i),b(i)\}}}
\end{equation}

The method gives similar speaker confusion errors to the traditional method with manually tuned threshold for each dataset, while not requiring such parameter optimisation. Note that the method cannot predict cluster of $1$ as the output.


\section{Experiments}
\label{sec:exp}

Experiments are performed to evaluate the proposed SAD method on three independent datasets -- our internal dataset of real-world conversations recorded with a single-channel microphone, the VoxConverse dataset containing videos of discussions from YouTube and the DIHARD challenge dataset.

The first two subsections describe the evaluation protocol and the baselines which are common across the three experiments, and the subsequent subsections will describe the experiments on each dataset.


\subsection{Evaluation protocol}

We use the Diarisation Error Rate (DER) as the overall performance metric. The DER is the sum of three error components: missed speech (MS, speaker in reference, but not in prediction), false alarm (FA, speaker in prediction, but not in reference) and speaker error (SC, assigned to wrong speaker ID). 

The tool used for evaluating the performance is the one developed for the rich transcription diarisation evaluations by NIST~\cite{istrate2005nist}. We include acceptance margin (collar) of 250 ms to compensate for human errors in ground truth annotation, except in the DIHARD datasets where the collar is zero.

\subsection{Baselines}

The baseline used in the ablation studies is divided into two strands: (1) full diarisation pipelines, (2) our pipeline with different speech activity detectors.

For the first case, we use a popular pipeline which is based on the winning entries on the 2018 DIHARD challenge~\cite{sell2018diarization,sun2018speaker}. This pipeline consists of a publicly available SAD system~\cite{johnston2012webrtc}, together with an x-vector based speaker embedding network followed by AHC. An optional speech enhancement module is used on the input. The parameters have been tuned on the DIHARD development set by the authors, and we do not tune or modify parameters. The recipes without and with speech enhancement are referred to as {\bf Baseline} and {\bf Baseline w/ SE} in the results.

For the second case, we follow the pipeline described in Section~\ref{sec:methods}, but replace the speech activity detector with off-the-shelf and pre-trained detectors. In particular, we use three baseline SAD systems -- an energy based detector and a pre-trained DNN-based detector both from the Kaldi toolbox~\cite{povey2011kaldi}\footnote{Using the Bob-Kaldi implementation~\cite{cernak2017bob}}, and the popular WebRTC detector~\cite{johnston2012webrtc}. Each of these SAD methods requires a threshold, which is set on each dataset. In the results tables, these methods are referred to as {\bf Ours w/ WebRTC SAD} and so on.


\subsection{VoxConverse}
\label{subsec:voxconverse}
The dataset includes videos from a range of multi-speaker acoustic environments, including political debates, panel discussions, celebrity interviews, comedy news segments and talk shows. This provides a number of background degradations, including dynamic environmental noise with some speech-like characteristics, such as laughter and applause.

The development set of \texttt{VoxConverse} consists of 216 multispeaker videos covering approximately 20 hours.
Table~\ref{table:voxconverse} summarises the dataset statistics. 

\begin{table}[h!]
\centering
\renewcommand\arraystretch{1.2}
\caption{\texttt{VoxConverse} dataset statistics. We use the dataset version {\tt v0.0.1}. \textbf{\# videos}: Number of sessions. \textbf{\# mins}: Total number of minutes in the dataset. \textbf{\# spks}: Min/mean/max values of unique speakers for each video. }
\label{table:voxconverse}
\tablespace
\begin{tabular}{ l  r  r r }
\toprule
 \textbf{set} & \textbf{\# videos} & \textbf{\# mins}  & \textbf{\# spks} \\ 
\midrule
Dev   & 216   & 1,212 & 1 / 4.5 / 20  \\ 
 \bottomrule
\end{tabular} 
\normalsize
\end{table}

Since this is an audio-visual dataset, we perform experiments using the audio-visual method of~\cite{chung2020spot} as well as the audio-only method. 
The audio-visual method is based on a combination of an active speaker detection method~\cite{Chung16a} and speaker verification using self-enrollment~\cite{chung2019said}, and will not be described here in detail.
The implementation details are identical to that reported in the original paper including the speaker embedding model, except for only the speech activity detector. Therefore in the audio-visual experiment, we use different speaker embedding extractors for the speaker representations and for the speech activity detection.

\begin{table}[h]
\centering
\renewcommand\arraystretch{1.2}
\setlength{\tabcolsep}{4pt}
\caption{
Results on the VoxConverse dataset using baseline methods and our proposed methods.
All values are in \%.
{\bf MS:} missed speech;
{\bf FA:} false alarm;
{\bf SC:} speaker confusion;
{\bf DER:} diarsation error rate (where $DER = MS + FA + SC$).
{\bf SE:} speech enhancement.
Note that the dataset version {\tt v0.0.1} is used.
$\dag$: reproduced using the code released in~\cite{dihard2019baseline}.
}
\label{table:results_vox}
\tablespace
\begin{tabular}{ l |  r  r  r r  }
\toprule
 \textbf{Name}   & \textbf{MS} & \textbf{FA}  & \textbf{SC} & \textbf{DER} \\ 
\midrule
Baseline based on~\cite{sell2018diarization} $\dag$                     & 11.1 & 1.6 & 11.1 & 23.7  \\ 
Baseline w/ SE based on~\cite{sell2018diarization,sun2018speaker} $\dag$ & 9.2 & 1.4 & 9.4 & 20.1   \\ 
\midrule 
Ours w/ Kaldi Energy SAD & 2.3 & 4.3 & 7.4 & 14.2 \\
Ours w/ Kaldi DNN SAD & 3.7 & 3.9 & 11.6 & 19.3 \\
Ours w/ WebRTC SAD & 5.2 & 2.4 & 7.2 & 14.8 \\
Ours w/ WebRTC SAD + SE & 4.5 & 2.3 & 6.2 & 13.0 \\
Ours w/ SpeakerNet SAD GMM & 4.0 & 0.7 & 4.2 & 9.0 \\
Ours w/ SpeakerNet SAD Fixed & 3.1 & 0.7 & 4.3 & 8.1 \\
\midrule 
AVSD w/ Kaldi Energy SAD & 2.9 & 3.4 & 3.2 & 9.5 \\
AVSD w/ Kaldi DNN SAD & 3.9 & 2.5 & 3.1 & 9.4 \\
AVSD w/ WebRTC SAD	   & 2.4 & 2.3 & 3.0 & 7.7	    \\
AVSD w/ SpeakerNet SAD GMM	   & 2.7 & 1.4 & 2.9 & 6.9	    \\
AVSD w/ SpeakerNet SAD Fixed	   & 2.3 & 1.4 & 3.0 & 6.7	    \\
 \bottomrule
\end{tabular} 
\end{table}

The results are reported in Table~\ref{table:results_vox}. Our extensive experiments demonstrate that the proposed method outperforms a range of baselines by a significant margin. The method is effective for audio-visual diarisation as well as audio-only. 

It is also notable that the use of SpeakerNet SAD reduces the speaker confusion error as well as the speech activity detection error. This reduction in error can be attributed to the fact that the proposed SAD utilises the speaker embedding confidence to reject less confident segments as non-speech. Therefore, the missed speech by SpeakerNet SAD is more likely contain the segments that would have nevertheless resulted in speaker confusion error.


\subsection{Internal conversations dataset}

This dataset covers a variety of recordings from informal discussions, offline meetings and zoom meetings. The recording equipment used in the experiment range from mobile phones to video conferencing microphone arrays, simulating ‘real-world' conditions. The recordings have been labelled professionally by trained annotators, with the aid of video. Table~\ref{table:internal} gives an overview of the dataset.

\begin{table}[h!]
\centering
\renewcommand\arraystretch{1.2}
\caption{Internal conversations dataset statistics. \textbf{\# mins}: Total number of minutes in the dataset. \textbf{\# spks}: Min/mean/max values of unique speakers for each video.}
\label{table:internal}
\tablespace
\begin{tabular}{ l  r  r r }
\toprule
 \textbf{set} & \textbf{\# sessions} & \textbf{\# mins}  & \textbf{\# spks} \\ 
\midrule
Dev   & 13   & 430 & 4 / 7.3 / 11  \\ 
 \bottomrule
\end{tabular} 
\normalsize
\end{table}

\begin{table}[h]
\centering
\renewcommand\arraystretch{1.2}
\setlength{\tabcolsep}{4pt}
\caption{
Results on the internal conversations dataset using the baseline and proposed methods.
All values are in \%.
The abbreviations are the same as in Table~\ref{table:results_vox}.
Note that the dataset version {\tt v0.0.1} is used.
$\dag$: reproduced using the code released in~\cite{dihard2019baseline}.
}
\label{table:results_internal}
\tablespace
\begin{tabular}{ l |  r  r  r r  }
\toprule
 \textbf{Name}   & \textbf{MS} & \textbf{FA}  & \textbf{SC} & \textbf{DER} \\ 
\midrule
Baseline based on~\cite{sell2018diarization} $\dag$                     & 26.4 & 3.1 & 16.4 & 45.9  \\ 
Baseline w/ SE based on~\cite{sell2018diarization,sun2018speaker} $\dag$ & 18.1 & 3.3 & 19.8 & 41.2   \\ 
\midrule
Ours w/ WebRTC SAD & 22.1 & 1.7 & 17.2 & 41.0 \\
Ours w/ WebRTC SAD + SE & 10.5 & 2.8 & 24.5 & 37.8 \\
Ours w/ SpeakerNet SAD GMM & 6.9 & 2.2 &  20.9 & 30.0 \\
Ours w/ SpeakerNet SAD Fixed & 10.3 & 1.6 & 22.0 & 34.0 \\
 \bottomrule
\end{tabular} 
\end{table}

Table~\ref{table:results_internal} gives the results on this dataset. The results are consistent with that in the VoxConverse dataset.


\subsection{DIHARD challenge dataset}
DIHARD is a series of challenges focusing on ‘hard' diarisation, where the state-of-the-art systems fare poorly. The data includes clinical interview, child language acquisition recordings, restaurant recordings and so on. 

In particular, we perform experiments on the development and test sets of the 2018 challenge data, each of which contains around 20 hours of speech data. The dataset statistics is given in Table~\ref{table:stats_dihard}.

\begin{table}[h!]
\centering
\renewcommand\arraystretch{1.2}
\caption{The DIHARD 2018 dataset statistics. \textbf{\# mins}: : Total number of minutes in the dataset. \textbf{\# spks}: Min/mean/max values of unique speakers for each video. }
\label{table:stats_dihard}
\tablespace
\begin{tabular}{ l  r  r r }
\toprule
 \textbf{set} & \textbf{\# sessions} & \textbf{\# mins}  & \textbf{\# spks} \\ 
\midrule
Dev   & 164   & 1,147 & 1 / 3.3 / 10  \\ 
Test   &  172   & 1,213 & 1 / 3.4 / 9  \\ 
 \bottomrule
\end{tabular} 
\normalsize
\end{table}

We compare our method to a range of baselines, including the ones that have been released specifically for this dataset. The baseline methods~\cite{sell2018diarization,sun2018speaker} are based on the winning submissions to the 2018 DIHARD challenge. 
\begin{table}[h]
\centering
\renewcommand\arraystretch{1.2}
\setlength{\tabcolsep}{4pt}
\caption{
Results on the DIHARD 2018 dataset using the baseline and proposed methods.
All values are in \%.
The abbreviations are the same as in Table~\ref{table:results_vox}.
$\dag$: reproduced using the code released in~\cite{dihard2019baseline}.
}
\label{table:results_dihard}
\tablespace
\begin{tabular}{ l |  r  r  r r  }
\toprule
 \textbf{Name}   & \textbf{MS} & \textbf{FA}  & \textbf{SC} & \textbf{DER} \\ 
\midrule
\multicolumn{5} {c} {\bf Development set} \\ 
Baseline based on~\cite{sell2018diarization} $\dag$                     &  21.2 & 7.5 & 11.4 & 40.1  \\ 
Baseline w/ SE based on~\cite{sell2018diarization,sun2018speaker} $\dag$ & 15.7 & 7.6 & 12.1 & 35.4   \\ 
\midrule
Ours w/ WebRTC SAD & 17.9 & 11.3 & 12.7 & 41.9 \\
Ours w/ WebRTC SAD + SE & 12.8 & 11.6 & 12.3 & 36.7 \\
Ours w/ SpeakerNet SAD GMM & 19.7 & 6.3 & 8.0 & 34.0 \\
Ours w/ SpeakerNet SAD Fixed & 20.2 & 7.1 & 7.6 & 35.0 \\
\midrule
\multicolumn{5} {c} {\bf Test set} \\ 
Baseline based on~\cite{sell2018diarization} $\dag$                     & 29.3 & 7.7 & 13.1 & 50.1  \\ 
Baseline w/ SE based on~\cite{sell2018diarization,sun2018speaker} $\dag$ & 18.4 & 7.8 & 15.9 & 42.1   \\ 
\midrule
Ours w/ WebRTC SAD & 26.4 & 10.9 & 13.4 & 50.6 \\
Ours w/ WebRTC SAD + SE & 15.4 & 11.7 & 16.3 & 43.4 \\
Ours w/ SpeakerNet SAD GMM & 22.3 & 6.2 & 11.0 & 39.5 \\
Ours w/ SpeakerNet SAD Fixed & 23.2 & 6.6 & 10.2 & 40.0 \\
 \bottomrule
\end{tabular} 
\end{table}

The results are reported in Table~\ref{table:results_dihard}, where our method consistently outperforms all baselines with comparable setup. The reduction in speaker confusion error (discussed in Section~\ref{subsec:voxconverse}) is also observed here.

While the winners of the challenge report slightly better DER figures compared to our systems, \cite{sell2018diarization} uses the Variation Bayes refinement and \cite{sun2018speaker} uses the Viterbi decoding-based refinement which are not used in our system or the baselines released in~\cite{dihard2019baseline}.


\section{Conclusion}

We can give a qualified answer to the question posed in the introduction: ‘‘Yes, it is possible to obtain speech activity labels directly from speaker embeddings, and to a surprisingly high standard''. We have proposed a simple but highly effective solution for speech activity detection based on the speaker embeddings. The method is applied to various diarisation experiments, in which we demonstrate competitive performance against all publicly available speech activity detectors. The code and the pre-trained models for this work will be released to the public.


\newpara{Acknowledgments.}
We would like to thank Icksang Han for helpful discussions.

\clearpage
\bibliographystyle{IEEEbib}
\bibliography{shortstrings,refs,vgg_local,vgg_other}

\end{document}